\documentclass[aps,twocolumn,showpacs]{revtex4}
\usepackage{graphicx}

\begin{document}

\title{Kinematical bound in asymptotically translationally invariant spacetimes}

\author{Tetsuya Shiromizu$^{(1,2)}$, Daisuke Ida$^{(1)}$ and Shinya Tomizawa$^{(1)}$}

\affiliation{$^{(1)}$Department of Physics, Tokyo Institute of Technology, 
Tokyo 152-8551, Japan}

\affiliation{$^{(2)}$Advanced Research Institute for Science and Engineering, 
Waseda University, Tokyo 169-8555, Japan}

\date{\today}

\begin{abstract}
We present positive energy theorems in asymptotically translationally invariant 
spacetimes which can be 
applicable to black strings and charged branes. We also address the bound property 
of the tension and charge of branes. 
\end{abstract}


\maketitle

\label{sec:intro}
\section{Introduction}

Asymptotic flatness is a useful working assumption in studying four-dimensional 
general relativity, in particular in the field of the black hole physics.
Of course, this assumption is well justified for the gravitational phenomena of isolated systems within the Hubble horizon of our universe.
Within the framework of the string theory, on the other hand, 
we have to take account of non-asymptotically
flat space-times, because the vacuum of the theory is considered to be four-dimensional
space-time times compact extra dimensions at low energy to realize our apparently four-dimensional universe. 
For example, we need more insight into the black strings of branes, which are non-asymptotically 
flat solutions typically arising in the supergravity theories, to extract some information on the
quantum gravity or the unified theory of interactions.

The stability of the Schwarzschild space-time is one of fundamental properties of the black holes in 
asymptotically flat space-time.
However, this  is not the case for the black strings or branes; 
namely, they are unstable under the linear perturbations of sufficiently long wave length
along the brane \cite{GL}.
We have no definitive answer concerning the end point of this Gregory-Laflamme instability,
but there are several possibilities; the final state might be naked singularities joining 
array of black holes, or inhomogeneous black string or brane \cite{Gary}, and there also be 
a possibility that there is no equilibrium state.
Since the subject concerns with non-asymptotically flat inhomogeneous
space-time, the analysis will be quite difficult. We would ultimately need dynamical analysis
directly solving the Einstein equation \cite{BS,Toby}.
It might be however also useful to have a kinematical bound irrelevant for the details of the
underlying theory for such non-asymptotically flat space-times.
Such a kinematical bound might be also useful to restrict the form of the metric 
like the uniqueness theorem in asymptotically flat spacetimes \cite{Israel,UniqTheo1}.

In this short report, utilizing
 the spinorial approach, we present bound theorems (positive mass
theorem, BPS bound, positive tension theorem) in asymptotically 
translationally invariant space-times. 
Recently Traschen 
discussed the positive mass theorem 
in such space-times without horizon and gauge fields\cite{Jennie2}. 
In this paper we will extend the Traschen's work to the cases with horizon in higher dimensions,
which is relevant for black string or brane space-times,
and include the gauge fields in four dimensions. 

The rest of the present paper is composed of two main parts. 
In Sec. II, we present the positive mass theorem in higher dimensions with horizon. Then 
we prove the positive energy and tension theorems for charged branes in Sec. III. 
Finally we 
 give a discussion in Sec. IV. In the appendix A we give formulae for the 
calculation of the boundary term at horizon.

\label{sec:ATI}

\section{Asymptotically translationally invariant space-times}

\begin{figure}[htbp]
\begin{center}
\includegraphics[width=.70\linewidth]{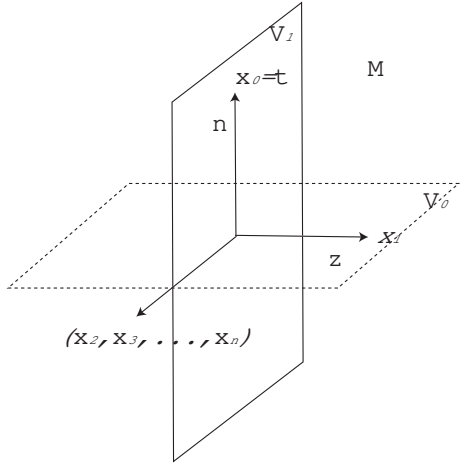}
\end{center}
\caption{Full space-time $\cal{M}$ can be foliated by 
spacelike hypersurfaces $V_0$ normal to timelike vector 
field $n\propto\partial_t$ and timelike hypersurfaces
 $V_1$ normal to spacelike vector field $z\propto\partial_{x^1}$. 
We can define coordinate $\{x^i\}=(x_1,x_2,\cdots,x_n)$ in $V_0$, 
$\{x^I\}=(x_0,x_2,\cdots,x_n)$ in $V_1$ and $\{x^A\}=(x_2,\cdots,x_n)$ 
in $(n-2)-$dimensional spacelike surface $V_{01}$ normal to 
both vector fields $n$ and $z$. Furthermore, we set 
coordinate $\{x^a\}=(x_3,\cdots,x_n)$ in $(n-3)$-dimensional spacelike 
submanifold $V_{012}$ normal to $n,z$ and $\hat{r}\propto\partial_{x^2}$.}
\label{fig1}
\end{figure}

First of all, we must specify asymptotically translationally invariant 
space-times. The metric of full space-times is given by 
%
\begin{eqnarray}
ds^2=g_{\mu\nu}dx^\mu dx^\nu. 
\end{eqnarray}
%
Let $n \propto \partial_t$ and $z \propto \partial_{x^1}$ to be timelike and spacelike unit 
vector fields such that $n^\mu z_\mu =0$. In addition, $\hat r \propto \partial_{x^2}$ to be spacelike 
perpendicular unit normal vector fields to $n$ and $z$, $\hat r^\mu n_\mu = \hat r^\mu z_\mu =0$. 
We assume that 
$z$ becomes to be proportional to asymptotically translational Killing vector toward the infinity 
directed to $\hat r$. $x^\mu$, $x^i$, $x^I$, $x^A$ and $x^a$ spans the full space-times ${\cal M}$, 
$(n-1)$-dimensional spacelike hypersurface $V_0$ normal to $n$, $(n-1)$-dimensional 
timelike hypersurface $V_1$ normal to $z$, 
$(n-2)$-dimensional spacelike submanifold $V_{01}$
orthogonal to $n$ and $z$, and $(n-3)$-dimensional spacelike submanifold $V_{012}$ orthogonal 
to $n$, $z$, $\hat r$. Each induced metricies can be written as 
%
\begin{eqnarray}
q_{ij}dx^i dx^j=(g_{\mu\nu}+n_\mu n_\nu ) dx^\mu dx^\nu
\end{eqnarray}
%
%
\begin{eqnarray}
h_{IJ}dx^I dx^J=(g_{\mu\nu}-z_\mu z_\nu ) dx^\mu dx^\nu
\end{eqnarray}
%
%
\begin{eqnarray}
p_{AB}dx^A dx^B=(g_{\mu\nu}+n_\mu n_\nu-z_\mu z_\nu) dx^\mu dx^\nu
\end{eqnarray}
%
and
%
\begin{eqnarray}
s_{ab}dx^a dx^b =(g_{\mu\nu}+n_\mu n_\nu -z_\mu z_\nu 
-\hat r_\mu \hat r_\nu) dx^\mu dx^\nu. 
\end{eqnarray}
%
Then $\mu,\nu=0,1,2,...,n$. $i=1,2,...,n$, $I=0,2,3,...,n$, 
$A=2,3,...n$, $a=3,4,...,n$. (See FIG.\ref{fig1}.)

We assume that the submanifold $V_{01}$ is $(n-2)$-dimensional asymptotically Euclid 
space. 

\label{sec:mass}

\section{Positive mass theorem for black string}

In this section, we present the positive energy theorem in asymptotically translationally 
invariant space-time with horizon. See Ref. \cite{Jennie2} for cases without horizon. 
If one thinks of the gravitational energy evaluated 
in slices which has appropriate asymptotic boundaries and regular center, it is not necessary to take 
the event horizon as the boundary term. However, the proof independent of the inner structure of horizon 
is useful. 

Let consider a spinor $\epsilon$ satisfying Dirac-type equation\cite{Jennie2}
%
\begin{eqnarray}
\gamma^{A} \nabla_A \epsilon =0. \label{WittenEq}
\end{eqnarray}
%
Note that we usually suppose $\gamma^i \nabla_i \epsilon =0 $ for spinor to prove 
the original positive energy theorem\cite{PET}. In asymptotically translationally invariant 
space-times, it is likely that the existence of solutions to Eq. (\ref{WittenEq}), 
which approaches constant spinor $\epsilon_0$, is guaranteed rather than the solution to 
$\gamma^i \nabla_i \epsilon=0$. This is because 
the space spanned by coordinate $ \lbrace x^A \rbrace $ is asymptotically flat 
and we can expect almost same proof of the existence of solutions 
with that in asymptotically flat space-times. 

Let us define the Nester tensor $E_{\mu\nu}$ by 
%
\begin{eqnarray}
E^{\mu\nu}=\frac{1}{2}\biggl( \bar \epsilon \gamma^{\mu\nu\alpha} 
\nabla_\alpha \epsilon +{\rm c.c.}\biggr) 
\end{eqnarray}
%
and we obtain formula 
%
\begin{eqnarray}
\nabla_\mu E^{\mu\nu}=\frac{1}{2}G^\nu_\mu \xi^\mu +{\overline {\nabla_\mu \epsilon}} \gamma^{\mu\nu\alpha}
\nabla_\alpha \epsilon\label{diverEq} 
\end{eqnarray}
%
where $\bar \epsilon = \epsilon^\dagger \gamma^{\hat 0}$. 
According to Ref. \cite{PET}, 
a surface integral of Nester tensor 
at spatial infinity over $V_{02}$ gives ADM energy-momentum vector $P^\mu$, that is,
\begin{equation}
-P^\mu \xi_\mu=\frac{1}{16\pi}\int_{V_{02}^{\infty}}E^{\mu\nu}dS_{\mu\nu}.
\end{equation} 
Integrating Eq.(\ref{diverEq}) over spacelike manifold $V_{02}$ 
and using Stokes's theorem and Eq.(\ref{WittenEq}), we obtain formula 
%
\begin{eqnarray}
& &  \int_{V_{02}^\infty}dS_{\hat0 \hat 2} E^{\hat 0 \hat 2} 
- \int_{V_{02}^{\rm H}}dS_{\hat0 \hat 2} E^{\hat 0 \hat 2} \nonumber \\
& & ~~~~= \int dV_0  \biggl( 8\pi T^\mu_\nu \xi^\nu n_\mu +2| \nabla_A \epsilon|^2 \biggr),
\end{eqnarray}
%
where $\xi^\mu = -\bar \epsilon \gamma^\mu \epsilon$. Following the proof in \cite{PET}, we require a spinor $\epsilon$ approaches a constant spinor $\epsilon_0$ at infinity $V_{02}^{\infty}$.  
In the above we used the Einstein equation $G_{\mu\nu}=8\pi T_{\mu\nu}$. 
The first and second terms in the left-hand side are boundary terms at 
infinity and horizon. The first term gives us the gravitational energy. 
Thus, what we must focus on is the boundary term at the horizon. This is 
non-trivial issue and the point here. We modify the proof in asymptotically 
flat space-times with horizon\cite{BH}. The detail of the computation is 
described in the appendix A. As a result, it becomes 
%
\begin{widetext}
\begin{eqnarray}
\int_{V_{02}^H}dS_{\hat0 \hat 2} E^{\hat 0 \hat 2} & = & 
\frac{1}{2}\int dS_{\hat 0 \hat 2} \Biggl[
\epsilon^\dagger ( \nabla_{\hat 2}
-\gamma^{\hat 2}\gamma^{\hat 1}\nabla_{\hat 1}  ) \epsilon
+{\rm c.c} \Biggr] \nonumber \\
& = & \frac{1}{2}\int dS_{\hat 0 \hat 2} 
\epsilon^\dagger \Biggl[ -\frac{1}{2}(K-K_{\hat 2 \hat 2}+k)\gamma^{\hat 2}
\gamma^{\hat 0} \epsilon - \gamma^{\hat 2} \gamma^{\hat 1} D_{\hat 1} \epsilon 
-\gamma^{\hat 2} \gamma^a d_{a} \epsilon +\frac{1}{2}K_{\hat a \hat 2}\gamma^{\hat a}
\gamma^{\hat 0}\epsilon \Biggr]+{\rm c.c.} \nonumber \\
& = &  \frac{1}{2}\int dS_{\hat 0 \hat 2} 
\epsilon^\dagger \Biggl[ -\frac{1}{2}(K-K_{\hat 2 \hat 2}+k)\gamma^{\hat 2}
\gamma^{\hat 0} \epsilon + \psi \Biggr] +{\rm c.c.} \label{BT1}
\end{eqnarray}
\end{widetext}
%
where 
%
\begin{eqnarray}
\psi = - \gamma^{\hat 2} \gamma^{\hat 1} D_{\hat 1} \epsilon 
-\gamma^{\hat 2} \gamma^a d_{a} \epsilon +\frac{1}{2}K_{\hat a \hat 2}\gamma^{\hat a}
\gamma^{\hat 0}\epsilon. 
\end{eqnarray}
%
$D_i, {\cal D}_A$ and $d_a$ are covariant derivative with respect to 
$q_{ij}$, $p_{AB}$ and $s_{ab}$, respectively. $K_{ij}$ and $k_{ab}$ 
are defined by $K_{ij}=q_i^k \nabla_k n_j $ and $k_{ab}=s_a^c {\cal D}_c \hat r_b$, 
respectively. 

At the horizon we impose 
%
\begin{eqnarray}
\gamma^{\hat 2} \gamma^{\hat 0} \epsilon = \epsilon
\end{eqnarray}
%
and use $\theta_+ \propto K-K_{\hat 2 \hat 2}+k=0$ at the apparent horizon. 
$\theta_+$ is the expansion of outgoing null geodesic congruence. 
Then we can see the boundary term at the horizon vanishes. We used the fact that 
$\psi$ anti-commutes with $\gamma^{\hat 2} \gamma^{\hat 0}$ and then 
the contribution of $\psi$ to Eq. (\ref{BT1}) disappears. Finally 
%
\begin{eqnarray}
E_{\rm ADM} & =& \frac{1}{8\pi|\epsilon_0|^2} \int_{V_{02}^\infty}dS_{\hat0 \hat 2} E^{\hat 0 \hat 2} \nonumber \\
& = & \frac{1}{8\pi |\epsilon_0|^2} \int dV_0  \biggl( 8\pi T^\mu_\nu \xi^\nu n_\mu +2| \nabla_A \epsilon|^2 \biggr).
\end{eqnarray}
%
Together with the dominant energy condition, we can see that $E_{\rm ADM}$ is 
positive definite. 

Let us discuss $M=0$ cases. In this case, 
%
\begin{eqnarray}
\nabla_{\hat A} \epsilon=0
\end{eqnarray}
%
and then 
%
\begin{eqnarray}
{}^{(n)}R_{\hat A \hat B\mu\nu} \gamma^{\mu\nu} \epsilon=0. 
\end{eqnarray}
%
From the above we see 
%
\begin{eqnarray}
{}^{(n)}R_{\mu\nu \alpha \beta}=0.
\end{eqnarray}
%
This means that the space-time with zero energy is flat. Even for 
asymptotically translationally invariant space-times, the ground state is 
flat space-time.

\section{Bound theorems for charged branes in four dimensions}

\subsection{Positive energy theorem for charged brane}

In this subsection, we extend Traschen's study to cases with gauge field in four dimensions. 
It is easy to extend to higher dimensions following Ref. \cite{GHT}.  
For this we define the following covariant tensor motivated by N=2 supergravity\cite{N2PET}:
%
\begin{eqnarray}
\hat \nabla_\mu \epsilon = \nabla_\mu \epsilon +\frac{i}{4}F_{\alpha \beta}
\gamma^{\alpha \beta} \gamma_\mu \epsilon.  
\end{eqnarray}
%
Let us consider spinor $\epsilon$ satisfying 
%
\begin{eqnarray}
\gamma^A \hat \nabla_A \epsilon=0. 
\end{eqnarray}
%
The Nester tensor is defined by 
%
\begin{eqnarray}
\hat E^{\mu\nu} & = & \frac{1}{2}\biggl( \bar \epsilon \gamma^{\mu\nu\alpha} 
\hat \nabla_\alpha \epsilon +{\bf c.c.} \biggr) \nonumber \\
& = & E^{\mu\nu}-i \bar \epsilon (F^{\mu\nu}-\gamma_5 \tilde F^{\mu\nu})\epsilon
\end{eqnarray}
%
and we obtain the following formula
%
\begin{eqnarray}
\nabla_\mu \hat E^{\mu\nu} & = &   \frac{1}{2}G^\nu_{~\mu} \xi^\mu + 
{\overline {\hat \nabla_\mu \epsilon}} 
\gamma^{\mu \nu \alpha} \hat \nabla_\alpha \epsilon \nonumber \\
& & -i \bar \epsilon (\nabla_\mu F^{\mu\nu}-\gamma_5 \nabla_\nu \tilde F^{\mu\nu}) \epsilon \nonumber \\
& & + 4\pi \bar \epsilon T^{\mu \nu}(F)\gamma_\nu \epsilon 
\end{eqnarray}
%
where $\tilde F^{\mu\nu}=(1/2)\epsilon^{\mu\nu\alpha\beta}F_{\alpha\beta}$ and 
%
\begin{eqnarray}
T_{\mu\nu}(F)=\frac{1}{4\pi}
\Biggl( F_{\mu}^{~\alpha}F_{\nu\alpha}-\frac{1}{4}g_{\mu\nu}F^2 \Biggr).
\end{eqnarray}
%
Integrating over the spacelike hypersurface, we are resulted in 
%
\begin{widetext}
\begin{eqnarray}
8\pi \epsilon^\dagger_0 \Biggl[E_{\rm ADM }-i \gamma^{\hat 0}(Q_{\rm e}-\gamma_5 Q_{\rm m})\Biggr] \epsilon_0
& = & \int_{V_{02}^\infty} dS_{\mu\nu} \hat E^{\mu\nu} \nonumber \\
& = & \int_{V_0} d\Sigma 
\Biggl[ G^\mu_\nu \xi^\nu n_\mu +2|\hat \nabla_A \epsilon|^2
-2i \bar \epsilon (j^\mu_{\rm e} -\gamma_5j^\mu_{\rm m} ) \epsilon 
n_\mu - 8\pi T_{\mu\nu}(F)\xi^\mu n^\nu \Biggr] \nonumber \\
& = & 
\int_{V_0} d\Sigma 
\biggl( 8\pi T^\mu_\nu \xi^\nu n_\mu +2|\hat \nabla_A \epsilon|^2
-2i\bar \epsilon (j^\mu_{\rm e} -\gamma_5j^\mu_{\rm m} ) \epsilon   n_\mu \biggr),
\end{eqnarray}
\end{widetext}
%
where 
%
\begin{eqnarray}
Q_{\rm e}= \frac{1}{8\pi}\int dS_{\mu\nu} F^{\mu\nu}
\end{eqnarray}
%
%
\begin{eqnarray}
Q_{\rm m}= \frac{1}{8\pi}\int dS_{\mu\nu} \tilde F^{\mu\nu}
\end{eqnarray}
%
%
\begin{eqnarray}
j^\mu_{\rm e} =\nabla_\nu F^{\nu\mu} 
\end{eqnarray}
%
and 
%
\begin{eqnarray}
j^\mu_{\rm m} =\nabla_\nu \tilde F^{\nu\mu}. 
\end{eqnarray}
%
From first line to second one, we used 
the Einstein equation $G_{\mu\nu}=8\pi ( T_{\mu\nu}(F)+T_{\mu\nu})$. Using the 
above and the dominant energy condition, we can obtain the BPS bound 
%
\begin{eqnarray}
E_{\rm ADM} \geq {\sqrt {Q_{\rm e}^2 + Q_{\rm m}^2}}. 
\end{eqnarray}
%
As the inequality is saturated, $\hat \nabla_{\hat A} \epsilon=0$ holds. 
In general, $\hat \nabla_{\hat 0} \epsilon \neq 0$ and  $\hat \nabla_{\hat 1} 
\epsilon \neq 0$. Since $\hat \nabla_{\hat A} \epsilon$ can be regarded as 
a infinitesimal local supersymmetric transformation of the gravitino, 
it is well-known fact that a part of supersymmetry is broken. 

We note that the current BPS bound theorem is slightly different from 
that given in Ref. \cite{GHT}. 

\subsection{Positive tension theorem for charged branes}

Let discuss the issue on the positive tension theorem\cite{Jennie2} or 
BPS bound\cite{1st}. As is discussed in \cite{Jennie2}, 
we can expect the tension of a brane is a conserved charge associated 
with an asymptotic spatial translational Killing vector parallel  
to the brane as ADM energy is one associated with an asymptotic time translational Killing vector. In analogy with the construction of positive energy theorem, the Nester tensor is 
defined by 
%
\begin{eqnarray}
\hat B^{\mu\nu} & = & \frac{1}{2}
\biggl( \tilde \epsilon \gamma^{\mu\nu\alpha}\hat \nabla_\alpha \epsilon 
+{\rm c.c.} \biggr) \nonumber \\
& = & \frac{1}{2}
\biggl( \tilde \epsilon \gamma^{\mu\nu\alpha}\nabla_\alpha \epsilon 
+{\rm c.c.} \biggr)  \nonumber \\
& & -\frac{1}{2} \Biggl[ i\tilde \epsilon (F^{\mu\nu}-\gamma_5 \tilde F^{\mu\nu})
\epsilon + {\rm c.c.}  \Biggr] \nonumber \\
& = & \frac{1}{2}
\biggl( \tilde \epsilon \gamma^{\mu\nu\alpha}\nabla_\alpha \epsilon 
+{\rm c.c.} \biggr) =B^{\mu\nu}
\end{eqnarray}
%
where $\tilde \epsilon = \epsilon^\dagger \gamma^{\hat 1}$. The integration over time is 
taken to be finite interval $\Delta t$. 
We should note that time direction in the construction of the previous theorem is replaced with $x^1$ direction.  In similar way as previous section, 
we can easily show 
%
\begin{eqnarray}
8\pi \mu |\epsilon_0|^2 
& = &    \frac{1}{\Delta t} \int_{V_1} dt dS_{\hat A} B^{\hat A \hat 2} \nonumber \\
& =  & \frac{1}{\Delta t} \int dV_1 \biggl(2 |\nabla_A \epsilon|^2
-8\pi T^{\rm tot}_{\mu \hat 1}\tilde \xi^\mu \biggr).
\end{eqnarray}
%
where $\tilde \xi^\mu = \tilde \epsilon \gamma^\mu \epsilon$ and 
$T_{\mu\nu}^{\rm tot}=T_{\mu\nu}(F)+T_{\mu\nu}$. 
We followed the Traschen's definition 
of the tension. See Refs. \cite{Jennie1,Jennie2,1st} for the issue of the 
definition. 

Note that the gauge field does not contribute to the tension. 
Thus BPS bound cannot be proven although it has been argued 
in Ref. \cite{1st}. To prove that in general cases, we must 
improve the proof non-trivially. 

\section{Summary}

In this paper we proved several bound theorems in asymptotically translationally 
invariant space-times. More precisely we could prove the positive energy theorem 
for space-times with event horizon such as black strings. We also proved  
positive energy and tension theorem for charged brane configurations. 
For current definition of the tension, the gauge field does not contribute to 
the tension. 

The positive energy theorem for black string space-times might be able to get 
insight into issue on the final fate. We might be able to 
prove a sort of uniqueness theorem using the positive energy theorem. 
Indeed, in asymptotically flat space-times, the uniqueness theorem for 
static black holes can be proved in this line \cite{UniqTheo1}. 

\section*{Acknowledgments}

We would like to thank Norisuke Sakai for fruitful discussions. To complete this work, the
discussion during and after the YITP workshops YITP-W-01-15 and  YITP-W-02-19
were useful. The work of TS was supported by Grant-in-Aid for Scientific
Research from Ministry of Education, Science, Sports and Culture of 
Japan(No.13135208, No.14740155 and No.14102004). The work of DI was supported by JSPS.

\vskip 1cm 

\appendix


\section{boundary term at horizon}

Here we present some useful formulae. Using Dirac-Witten equation, 
$\nabla_{\hat 2}\epsilon$, which appeared as the first term in the integrand of 
the right-hand side in the first line of Eq. (\ref{BT1}), can be written as 
%
\begin{eqnarray}
\nabla_{\hat 2}\epsilon & = & -\gamma^{\hat 2} \gamma^{\hat a} \nabla_{\hat a} \epsilon 
\nonumber \\
& = & -\gamma^{\hat 2}\gamma^{\hat a} \biggl( 
D_{\hat a} \epsilon+\frac{1}{2}K_{\hat a \hat i} \gamma^{\hat i} \gamma^{\hat 0} \epsilon 
\biggr) \nonumber \\
& = & -\gamma^{\hat 2} \gamma^{\hat a} \biggl( {\cal D}_{\hat a} \epsilon
+\frac{1}{2}J_{\hat a \hat A} \gamma^{\hat A} \gamma^{\hat 1}\epsilon 
+\frac{1}{2}K_{\hat a \hat i} \gamma^{\hat i} \gamma^{\hat 0} \epsilon \biggr) \nonumber \\
& = & -\gamma^{\hat 2} \gamma^{\hat a} \biggl( d_{\hat a}\epsilon 
+\frac{1}{2}k_{\hat a \hat b} \gamma^{\hat b} \gamma^{\hat 2} \epsilon
+\frac{1}{2}J_{\hat a \hat A} \gamma^{\hat A} \gamma^{\hat 1}\epsilon \nonumber \\
& & ~~+\frac{1}{2}K_{\hat a \hat i} \gamma^{\hat i} \gamma^{\hat 0} \epsilon 
  \biggr) \nonumber \\
& = & -\gamma^{\hat 2} \gamma^{\hat a} d_{\hat a}\epsilon -\frac{1}{2}k \epsilon
-\frac{1}{2}J_{\hat a \hat A} \gamma^{\hat 2} \gamma_{\hat a} \gamma^{\hat A} \gamma^{\hat 1} \epsilon 
\nonumber \\
& & -\frac{1}{2}K_{\hat a \hat i} \gamma^{\hat 2} \gamma^{\hat a} \gamma^{\hat i} \gamma^{\hat 0} \epsilon
\end{eqnarray}
%
where $J_{AB}$ is defined by $J_{AB}=p_A^C D_C z_B$. 

Let us define a scalar field $\phi$ by 
%
\begin{eqnarray}
\phi := \epsilon^\dagger J_{\hat a \hat A} \gamma^{\hat 2} \gamma^{\hat a} 
\gamma^{\hat A} \gamma^{\hat 1} \epsilon 
= - \epsilon^\dagger J_{\hat a \hat 2} \gamma^{\hat 2} \gamma^{\hat a} \gamma^{\hat 1} \epsilon
+\epsilon^\dagger J_{\hat a}^{~\hat a} \epsilon^\dagger \gamma^{\hat 2} \gamma^{\hat 1} \epsilon.
\end{eqnarray}
%
It is easy to see that $\phi$ is pure imaginal, $\phi^*=-\phi$. Then
%
\begin{eqnarray}
{\rm Re} (\epsilon^\dagger  \nabla_{\hat 2} \epsilon)  & = &  
-\epsilon^\dagger \gamma^{\hat 2} \gamma^{\hat a} d_{\hat a}\epsilon
-\frac{1}{2}k |\epsilon|^2+\frac{1}{2}K_{\hat a \hat 2} \epsilon^\dagger \gamma^{\hat a} 
\gamma^{\hat 0} \epsilon \nonumber \\
& & - \frac{1}{2}K_{\hat a}^{\hat a} \epsilon^\dagger \gamma^{\hat 2} \gamma^{\hat 0} \epsilon
\end{eqnarray}
%
In a same way, we obtain the following formula for the 
second term of the integrand in the right-hand side in the 1st line 
of Eq. (\ref{BT1}): 
%
\begin{eqnarray}
{\rm Re} (\epsilon^\dagger  \gamma^{\hat 2} \gamma^{\hat 1}\nabla_{\hat 1} \epsilon) 
& = &  {\rm Re}\Biggl[ \epsilon^\dagger  \gamma^{\hat 2} \gamma^{\hat 1} 
\biggl( D_{\hat 1} \epsilon +\frac{1}{2}K_{\hat 1 \hat i} \gamma^{\hat i} \gamma^{\hat 0}\epsilon  
\biggr) \Biggr] \nonumber \\
& = &   {\rm Re} \Biggl[\epsilon^\dagger  \gamma^{\hat 2} \gamma^{\hat 1}  D_{\hat 1} \epsilon \Biggr] 
\nonumber \\
& & + \frac{1}{2}K_{\hat 1 \hat 1}{\rm Re} \Biggr[ \epsilon^\dagger \gamma^{\hat 2} \gamma^{\hat 0}\epsilon
\Biggr]
\end{eqnarray}
%

\end{document}